\documentstyle[12pt]{report}
\begin{document}
\baselineskip=20pt

\begin{center}
{\bf ON THE NATURE OF THE GAUSSIAN APPROXIMATIONS}\\
{\bf IN PHASE ORDERING KINETICS}\\
{}~~\\
{}~~\\
{}~~\\
S. De Siena and M. Zannetti\\
{\small \it Dipartimento di Fisica, Universit\`{a} di Salerno\\
84081 Baronissi (SA), Italy}
\end{center}

{}~~\\

\begin{center}
{\bf Abstract}
\end{center}

\begin{quote}
The structure of the gaussian auxiliary field approximation
in the theory of phase ordering kinetics is analysed with
the aim of placing the method within the context of a systematic
theory. While we are unable to do this for systems with a scalar
order parameter, where the approximation remains uncontrolled,
a systematic development about the gaussian approximation
can be outlined for systems with a vector order parameter
in terms of a suitably defined $1/N$-expansion.
\end{quote}

{}~~\\
05.70.Fh 64.60.Cn 64.60.My 64.75.+g

\newpage

\setcounter{chapter}{1}

\section*{1 - Introduction}

\vspace{5mm}

Although much progress has been made in the understanding of
phase ordering kinetics$^{1,2}$, from the point of view of the theorist
basically this remains an unsolved problem. The reason is that
in the case of a scalar order parameter, which is the most
relevant for experiments, no systematic scheme for the development
of a perturbation theory is available. This requires the existence
of a soluble zero order approximation which accounts, at least
qualitatively, for the relevant physical features of the
problem and of a well defined procedure for the calculation, at
least in principle, of the higher order corrections. The case
of a system with continous symmetry is in better shape since
the $1/N$-expansion meets
these requirements, at least in the case of a non conserved order
parameter. For a conserved order parameter there are
indications that the large-N limit might be singular$^{3}$.

Despite this very unsatisfactory situation, recently much progress
has been made in the development of analytical methods for the computation
of the structure factor, through extensive use of approaches based on the
introduction of a gaussian auxiliary field$^{4,5}$ (GAF) which improve on
the original idea of Ohta, Jasnow and Kawasaki$^{6}$ (OJK).
An exaustive critical
account of these theories has been given by Yeung, Oono
and Shinozaki$^{5}$.
The success of this approach, which for the moment is mostly limited
to non conserved order parameter, amounts to the very accurate
reproduction of the scaling function for scalar order parameter$^{4,5,6}$
as known from experiments or numerical simulations, and to the prediction
of power law tails in the case of vector order parameter$^{7,8}$.
Thus, these theories seem to incorporate those basic ingredients that a real
theory of phase ordering kinetics should have. The shortcoming is that
the assumption that the auxiliary field obeys gaussian statistics
is totally uncontrolled. More than theories for the moment
these are sophisticated computational prescriptions which
are justified a posteriori.

A substantial progress toward a systematic theory then would be made
if it were possible to identify a scheme within which a GAF approximation
plays the role of the zero order approximation together with the expansion
parameter which generates the higher order corrections. This type of project
is illustrated in the paper of Bray and Humayun$^{9}$.

Motivated by these considerations here we overview the GAF approximations
with the aim of exposing those features which help to put into focus
what is required for the eventual development of a systematic theory.
For pedagogical reasons we begin with a detailed discussion of the one
particle problem which is exactly soluble and therefore allows to
illustrate clearly what is involved in a GAF-type approximation.
The same pattern of analysis than will be applied to the field theory
case.

\vspace{8mm}

\setcounter{chapter}{2}
\setcounter{equation}{0}

\section*{2 - One Particle}

\vspace{5mm}

Let us consider one particle in a double well potential and in contact
with a thermal reservoir. The decay process from the instability point
of this system has been thouroughly studied in the literature$^{10}$.
In the limit of zero temperature the equation of motion for the position
$\phi(t)$ is given by
\begin{equation}
{\dot \phi} = r\phi-g\phi^3
\end{equation}
with $r>0 \,,\, g>0$. In order to study the quench of this system
from high temperature to zero temperature, let us consider a gaussian
probability distribution for the initial condition $\phi_{0}=\phi(t_{0})$
\begin{equation}
P_{0}(\phi_{0}) = {1 \over {\sqrt{2 \pi \Delta}}}
e^{-{\phi_{0}^2 \over {2 \Delta}}}.
\end{equation}
Due to the symmetry of the problem the average position of the particle
vanishes identically $<\phi(t)> \equiv 0$ and
we concentrate on the behaviour of the fluctuations
\begin{equation}
S(t) = <\phi^2(t)> = \int_{-\infty}^{+\infty} d\phi P(\phi,t) \phi^2
\end{equation}
where $P(\phi,t)$ is the probability that the particle occupies the
position $\phi$ at the time $t$. This quantity can be computed exactly
since the equation of motion (2.1) can be solved
\begin{equation}
\phi(t) = f(t-t_{0},\phi_{0}) = {{\tau \phi_{0}}
\over {[1+(g/r)\phi_{0}^2(\tau^2-1)]^{1/2}}}
\end{equation}
with $\tau=e^{r(t-t_{0})}$. Thus, for the probability density we find
\begin{eqnarray}
&& P(\phi,t)=P_{0} \bigl(f^{-1}(t-t_{0},\phi) \bigr) {{df^{-1}(t-t_{0},\phi)}
\over {d\phi}} \nonumber \\
&& \\
&&={1 \over {\sqrt{2 \pi \Delta \tau^2}}} {{\exp \{-{1 \over {2 \Delta \tau^2}}
{\phi^2 \over {[1-(g/r)\phi^2(1-\tau^{-2})]}}\} } \over
{[1-(g/r) \phi^2 (1-\tau^{-2})]^{3/2}}} \nonumber
\end{eqnarray}
where $f^{-1}(t-t_{0},\phi_{0})$ is the inverse of
$f(t-t_{0},\phi_{0})$. Inserting into (2.3)
\begin{equation}
S(t)=\frac{r}{g}\frac{\tau^2}{(\tau^2-1)} \bigl\{1-{\sqrt{\pi x}}e^{x}
[1-erf({\sqrt{x}})] \bigr\}
\end{equation}
where $x=r/[2 \Delta g (\tau^2-1)]$ and $erf(z)$ is the error function.
For short time (2.6) yields exponential growth of the fluctuations
\begin{equation}
S(t) \sim \Delta \tau^2
\end{equation}
due to the initial instability, while for large time we get
\begin{equation}
S(t) \sim \frac{r}{g}\frac{\tau^2}{(\tau^2-1)}[1-{\sqrt{\pi x}}]
\end{equation}
which describes the saturation toward the finite equilibrium value
$S(\infty)=\phi^{2}_{eq}=r/g$ due to the fact that eventually the particle sits
at the bottom of one of the two potential wells and the probability density
(2.5) develops two narrow peaks centered about the equilibrium values
$\phi_{eq}=\pm \sqrt{r/g}$.

If the exact solution of the problem was
not available, this type of behaviour could not have been obtained
via a straightforward perturbation expansion in the nonlinear coupling $g$.
Zero order amounts to take the gaussian approximation in (2.5)
which describes only the regime of exponential growth (2.7).
Hence, the zero order theory does not reproduce the  qualitative
picture of the process, neither any improvement is obtained by taking
into account corrections of finite order.
The saturation to a finite final equilibrium value
is obtained  within a perturbative scheme, as shown
by Suzuki$^{10}$, by resorting  to the
infinite resummation of the most divergent terms in the series.

However, rather than following this route, let us use the method
which in the following will be generalized to the field theory case.
The idea is to introduce a new auxiliary
variable $m(t)$ through a transformation
\begin{equation}
\phi(t)=\sigma(m(t))
\end{equation}
which takes care of
the basic non-linear features of the problem
in such a way that the behaviour of $m(t)$ can be treated by
straightforward perturbation theory. Namely, the transformation must be
such that while $m(t)$ is allowed to grow indefinitely
the saturation of $\phi(t)$ to a finite value is induced
by $\sigma$.

Substituting (2.9) into (2.1) one obtains the equation of motion
for $m(t)$
\begin{equation}
{\dot m}={\sigma(m) \over \sigma^{'}(m)}[r-g \sigma^2 (m)]
\end{equation}
with the transformed probability density of initial conditions
\begin{equation}
P_{m_{0}}(m_{0})=P_{0}(\sigma(m_{0})|R) \sigma^{\prime}(m_{0})
\end{equation}
where $m_{0}=m(t_{0})$ and $P_{0}(\phi_{0}|R)$
is the probability density (2.2) conditioned
to $\phi_{0}$ belonging to the range of values $R$ for which (2.9) is
invertible.
Thus, in terms of $m(t)$ we cannot quite get  the fluctuations (2.3),
but fluctuations conditioned to $\phi \in R$
\begin{equation}
S(t|R)=\int d\phi P(\phi,t|R) \phi^{2}=\int dm P_{m}(m,t) \sigma^{2}(m)
\end{equation}
where $R$ is the domain
\begin{equation}
R=(\phi^{2} \leq r/g)
\end{equation}
and $P_{m}(m,t)$ is
the probability density of $m$ at the time $t$. How important this
restriction is depends on what is the statistical weight of trajectories
lying outside $R$ and this in turn is related to the size of the
variance $\Delta$ of the initial probability density (2.2)
compared to the size $r/g$ of the domain $R$. In the following we
shall neglect the distinction between $S(t|R)$ and $S(t)$ by
assuming $\Delta \ll r/g$.

Now, if the transformation $\sigma$ is such that (2.10) can be
solved, at least in perturbation theory, denoting the solution by
$m(t)=h(t-t_{0},m_{0})$ we have
\begin{eqnarray}
&& P_{m}(m,t)=P_{m_{0}} \bigl(h^{-1}(t-t_{0},m)\bigr)
{{dh^{-1}(t-t_{0},m)} \over {dm}}= \nonumber \\
&& \\
&& P_{0} \bigl(\sigma\bigl(h^{-1}(t-t_{0},m) \bigr)|R \bigr)
{{dh^{-1}(t-t_{0},m)} \over {dm}}
\sigma^{\prime} \bigl(h^{-1}(t-t_{0},m) \bigr) \nonumber
\end{eqnarray}
which formally solves the problem since it gives an explicit
expression for $P_{m}(m,t)$ in terms of the known quantities
$P_{0}$,$\sigma$ and $h$.

In order to see how this works in practice let us go back to
the equation of motion (2.10) for $m(t)$ and let us look for the
transformation $\sigma$ which simplifies as much as possible the
behaviour of $m(t)$. The first attempt is for an outright
linearization of (2.10). If this was not possible then, as
stated above, $\sigma$ ought to be such that (2.10) can be solved
in perturbation theory. However, in this case linearization can
be achieved. Imposing
\begin{equation}
{\sigma(m) \over \sigma^{\prime}(m)}[r-g\sigma^{2}(m)]=rm
\end{equation}
one finds
\begin{equation}
\phi=\sigma(m)={m \over [1+(g/r)m^{2}]^{1 \over 2}}
\end{equation}
and
\begin{equation}
m(t)=h(t-t_{0},m_{0})=\tau m_{0}.
\end{equation}
Indeed, we have that while $m(t)$ grows exponentially
$\phi(t)$ eventually saturates via (2.16) to the final
equilibrium value $\phi_{eq}=\pm \sqrt{r/g}$. Namely, the
transformation $\sigma$ accounts for the nonperturbative
features of the problem.

Putting together (2.14),(2.16) and (2.17) we have the exact
solution of the problem in terms of the auxiliary variable
$m(t)$. The
motivation for going to this form of the solution is that in more
complicated cases where $h$ and therefore $P_{m}(m,t)$
cannot be explicitely obtained, the consideration that
the auxiliary variable $m(t)$
should not be much affected by the nonlinear nature of the problem
authorizes to attempt a gaussian ansatz for $P_{m}(m,t)$. This
will be the crucial step of the GAF approximation in the phase
ordering problem. The difficulty with an ansatz however is that
it may not be possible to control the corrections to it. In
any case it should be clear that a gaussian ansatz does not
amount to an overall linearization of the problem, since in
(2.12) the ansatz amounts to use a gaussian form for $P_{m}(m,t)$
while the nonlinearity remains through the explicit factor $\sigma^{2}(m)$.
To be more specific, since $P_{0}$ is gaussian it is evident from
(2.14) that $P_{m}(m,t)$ is gaussian if $\sigma$ and $h$
are linear. Thus, should it be possible to find an expansion
parameter $\lambda$ such that $\sigma$ and $h$ become linear for
$\lambda \rightarrow 0$, the gaussian approximation amounts
to take this limit inside $P_{m}(m,t)$ in (2.12) {\it but not}
in the explicit factor $\sigma^{2}(m)$.

Let us see how this works in the one particle context. Since in this
case $h(t-t_{0},m)$ is already linear, in order to make
$P_{m}(m,t)$ gaussian we need to linearize only $\sigma$ in (2.14).
{}From (2.16) we may write
\begin{equation}
\sigma^{2}(\tau^{-1}m)=\tau^{-2} m^{2}-\frac{g}{r}{{\tau^{-4}m^{4}}
\over {[1+(g/r)\tau^{-2}m^{2}]}}
\end{equation}
and using this in (2.14) we obtain
\begin{equation}
P_{m}(m,t)=P^{(0)}_{m}(m,t)K(\tau^{-1}m,g/r)
\end{equation}
with
\begin{equation}
P^{(0)}_{m}(m,t)={1 \over \sqrt{2 \pi \Delta \tau^{2}}}
\exp \{-{m^{2} \over {2 \Delta \tau^{2}}} \}
\end{equation}
and
\begin{equation}
K(\tau^{-1}m,g/r)=\frac{\exp \{\frac{g}{r}\frac{\tau^{-4}m^{4} }
{[1+(g/r)\tau^{-2}m^{2}]} \} }{[1+(g/r)\tau^{-2}m^{2}]^{\frac{3}{2}}}.
\end{equation}
Thus, in this case it is possible to identify the nonlinear
coupling $g$ with the expansion parameter $\lambda$ which generates
gaussian statistics for $m$ in the limit $\lambda \rightarrow 0$.
Then, following the previous discussion, the lowest order is
obtained by letting $g \rightarrow 0$ in (2.19) but not in
the explicit factor $\sigma^{2}(m)$. From (2.12) then we get
\begin{equation}
S^{(0)}(t)=\int_{-\infty}^{\infty}
dm P^{(0)}(m,t) \sigma^{2}(m)=
\frac{r}{g} \bigl\{1-\sqrt{\pi y}e^{y}[1-erf(\sqrt{y})] \bigr\}
\end{equation}
with $y=r/(2\Delta \tau^{2} g)$ which gives $S^{(0)}(t) \sim
\Delta \tau^{2}$ at short time as in (2.7) and $S^{(0)}(t)
\sim (r/g) [1-\sqrt{\pi y}]$ for long time. The
qualitative effect of the saturation is correctly reproduced,
altough there is a quantitative discrepancy with (2.8) in the
law of approach to equilibrium. In conclusion, in the one
particle case the gaussian approximation can be identified
with the zero order step in a systematic development where higher
order corrections are generated by expanding $K(\tau ^{-1}m,g/r)$
in powers of $g$.

\vspace{8mm}

\setcounter{chapter}{3}
\setcounter{equation}{0}

\section*{3 - Phase Ordering Dynamics}

\vspace{5mm}

Let us now turn to the field theory case. The phase ordering dynamics
following the quench from high temperature to zero temperature
of a system with a non conserved order parameter is described by
the equation of motion
\begin{equation}
\frac{\partial \phi(\vec x,t)}{\partial t}=\nabla^{2} \phi(\vec x,t)
-V^{\prime} \bigl(\phi(\vec x,t) \bigr)
\end{equation}
with a gaussian initial state which generalizes (2.2)
\begin{equation}
P_{0} \bigl[\phi_{0}(\vec x) \bigr]={1 \over {Z_{0}}}
e^{-{1 \over {2 \Delta}} \int d^{d}x \phi_{0}^{2}(\vec x)}
\end{equation}
and where $V(\phi)$ is a potential of the double well type.

Again, due to the symmetry of the problem the average order
parameter vanishes identically $<\phi(\vec x,t)> \equiv 0$
and the observable of interest is the equal time correlation
function
\begin{equation}
G(\vec u,t)=<\phi(\vec x_{1},t) \phi(\vec x_{2},t)>=
\int d\phi_{1} d\phi_{2} P(\phi_{1},{\vec x_{1}} t;\phi_{2},{\vec x_{2}} t)
\phi_{1} \phi_{2}
\end{equation}
or the structure factor
\begin{equation}
C(\vec{k},t)=\int d^{d}x e^{i\vec{k} \cdot \vec{u}}
G(\vec{u},t)
\end{equation}
where $\vec{u}=\vec{x}_{1}-\vec{x}_{2}$. In (3.3) $P(\phi_{1},\vec{x}_{1}t;
\phi_{2},\vec{x}_{2}t)$ is the joint probability density that
$\phi(\vec{x},t)$ takes the value $\phi_{1}$ at the space-time point
$(\vec{x}_{1},t)$ and the value $\phi_{2}$ at the space-time point
$(\vec{x}_{2},t)$.

It has been well established$^{1,2}$, both from experiment and numerical
simulations, that in the late stage of the dynamics these quantities
obey scaling
\begin{equation}
G(\vec{u},t) \sim f(u/L(t))
\end{equation}
\begin{equation}
C(\vec{k},t) \sim L^{d}(t)g(kL(t))
\end{equation}
where $L(t)$ is the basic length in the problem which
is related to the average size of domains and obeys the growth
law $L(t) \sim t^{1/2}$, while $f(x)$ and $g(x)$
are scaling functions. The origin of scaling is
that  in the late stage the order parameter
reaches local equilibrium and forms domains of the ordered
phases which evolve according to self-similar patterns.
{}From the existence of sharp interfaces separating domains
one can deduce$^{2}$ the short distance behaviour of $f(x)$ or
the long wavelength behaviour of $g(x)$ (Porod's law)
\begin{equation}
f(u/L)=1-2u/L+...\,\,\,\, for \,\, u/L<<1
\end{equation}
\begin{equation}
g(kL) \sim (kL)^{-(d+1)} \,\,\,\, for \,\, kL>>1
\end{equation}
as well as the saturation law$^{4}$ of the order parameter
\begin{equation}
S(t)= G(\vec{u}=0,t)=\phi_{eq}^{2} \bigl[1-\frac{a}{L(t)}+O(L^{-2}) \bigr]
\end{equation}
where $\phi_{eq}$ is the value of the order parameter in the final
equilibrium state. Eq.s from (3.5) to (3.9) contain the minimal
phenomenological information that a theory of phase ordering
dynamics should account for.

At this point it is important to emphasize that the scaling behaviour
described above applies to the late stage of the process where
domains are close to saturation and grow through the motion of the
interfaces. This stage of the dynamics is dominated by the nonlinear
nature of the problem and much as in the one particle case it cannot
be obtained through any straightforward perturbation expansion.
The great difference with the one particle case is that
Eq. (3.1) cannot be solved for any realistic potential. Therefore,
in order to make analytic progress, we turn to the generalization
of the auxiliary variable method.

\vspace{8mm}

\setcounter{chapter}{4}
\setcounter{equation}{0}

\section*{4 - Auxiliary Field Method}

\vspace{5mm}

Following the idea illustrated above we now introduce an auxiliary
field through a local nonlinear transformation
\begin{equation}
\phi (\vec x,t)=\sigma (m(\vec x,t))
\end{equation}
which in general is defined through a relation involving the
potential
\begin{equation}
K[\sigma(m)]=V^{\prime}(\sigma).
\end{equation}
We note that such a transformation cannot be a linearising transformation
as it was in the one particle case. In fact in that case $m(\vec x,t)$
ought to be the free field and the relation between the free field
and the interacting field is certainly nonlocal, as it can be easily
seen generating the formal solution of (3.1) by iteration.
 Thus, the transformation
(4.1) is introduced in order to take care at least of the gross
nonlinear effect which is the saturation of the order parameter
to the finite final equilibrium value $\phi_{eq}$, leaving the rest,
possibly, to perturbation theory. Accordingly,
for large time the transformation must go over to the form
\begin{equation}
\sigma (m(\vec x,t))=\phi_{eq} \,\, sign(m(\vec x,t)).
\end{equation}

The equation of motion of the auxiliary field is obtained from (3.1)
\begin{equation}
\frac{\partial m}{\partial t}=\nabla^{2}m+\frac{1}{\sigma^{\prime}}
[\sigma^{\prime \prime}(m)(\nabla m)^{2}-V^{\prime}(\sigma)]
\end{equation}
with the transformed initial condition
\begin{equation}
P_{m_{0}}[m_{0}(\vec x)]=P_{0}[\sigma(m_{0}(\vec x))]J(\phi_{0},m_{0})
\end{equation}
where $J(\phi_{0},m_{0})$ is the Jacobian of the transformation (4.1)
at the initial time. Representing the solution of (4.4) as a
functional of the initial configuration labeled by $\vec x$ and $t$
\begin{equation}
m(\vec x,t)=h(\vec x,t-t_{0};[m_{0}({\vec x}^{\prime})])
\end{equation}
the probability of a configuration $[m(\vec x)]$ at the time $t$
can be obtained in terms of the initial probability density (3.2)
\begin{eqnarray}
&& P_{m}[m(\vec x),t]=P_{m_{0}} \bigl[h^{-1}(\vec x,t-t_{0},
[m({\vec x}^{\prime})]) \bigr]J(m_{0},m) \nonumber \\
&& \\
&& =P_{0} \bigl[\sigma
\bigl(h^{-1}(\vec x,t-t_{0},[m({\vec x}^{\prime})]\bigr) \bigr) \bigr]
J(\phi_{0},m_{0})J(m_{0},m) \nonumber
\end{eqnarray}
where $h^{-1}$ is the inverse of (4.6) and $J(m_{0},m)$
is the Jacobian of this transformation.
The above result is the analogue of (2.14) and specifies the
statistics of the auxiliary field $m(\vec x,t)$ in terms of
$\sigma , h$ and the statistical properties of the initial condition.

Neglecting for simplicity considerations pertaining to the restriction
of averages to domains of configurations where (4.1) is invertible,
the correlation function (3.3) may be rewritten as
\begin{equation}
G(\vec u,t)=\int dm_{1} dm_{2}
P_{m}(m_{1},{\vec x}_{1} t;m_{2},{\vec x}_{2} t)
\sigma(m_{1}) \sigma(m_{2})
\end{equation}
where the joint probability of $m$ is related to (4.7) by
\begin{equation}
P_{m}(m_{1},{\vec x}_{1} t;m_{2},{\vec x}_{2} t)=
\int d[m(\vec x)] P_{m}[m(\vec x),t]
\delta(m_{1}-m({\vec x}_{1})) \delta (m_{2}-m({\vec x}_{2})).
\end{equation}
The above form (4.8) for the correlation function makes a progress
over (3.3) if the joint probability of $m$ is available.
This requires that the transformation $\sigma$ is such that Eq.(4.4)
for $m$ is soluble. Short of this, as explained in section 2, one resorts
to the GAF approximation through the linearization of $\sigma$
and $h$ inside $P_{0}$.

Let us now review the predictions of the GAF approximations.
If $m(\vec x,t)$ is gaussian the probability densities are of
the form
\begin{equation}
P_{m}^{(0)}(m_{1},{\vec x}_{1} t;m_{2},{\vec x}_{2} t)=
\frac{1}{Z_{m}} \exp \{-\frac{1}{2(1-\gamma^2)S_{0}(t)}
[m_{1}^{2}+m_{2}^{2}-2\gamma m_{1} m_{2}] \}
\end{equation}
and
\begin{equation}
P_{m}^{(0)}(m,{\vec x} t)=\frac{1}{\sqrt{2 \pi S_{0}(t)}}
\exp \{-\frac{m^{2}}{2S_{0}(t)} \}
\end{equation}
with
\begin{eqnarray}
S_{0}(t)=<m^{2}(\vec x,t)>_{0} \,\, & , &  \,\,
G_{0}(\vec u,t)=<m({\vec x}_{1},t)m({\vec x}_{2},t)>_{0}
\nonumber \\
& & \\
\gamma=\gamma(\vec u,t)=\frac{G_{0}(\vec u,t)}{S_{0}(t)}
\,\, & , & \,\,
Z_{m}=2 \pi S_{0}(t) \sqrt{1-\gamma^{2}} \nonumber
\end{eqnarray}
and where $<\cdot>_{0}$ denotes averages with respect to $P_{m}^{(0)}$.
Hence, for the fluctuations of the order parameter one has
\begin{equation}
S(t)=\int dm P_{m}^{(0)}(m, \vec x,t)\sigma^{2}(m)
\end{equation}
and for the scaling function
\begin{equation}
f \bigl(u/L(t) \bigr)=\int dm_{1} dm_{2}
P_{m}^{(0)}(m_{1},{\vec x}_{1} t;m_{2},{\vec x}_{2} t)
sign(m_{1}) sign(m_{2})=\frac{2}{\pi} {sin}^{-1}(\gamma).
\end{equation}
Within this approach the problem is reduced to the computation
of $G_{0}(\vec u,t)$.

For $h$ to be linear Eq. (4.4) must be of the form
\begin{equation}
\frac{\partial m}{\partial t}=\nabla^{2}m+a(t)m
\end{equation}
where $a(t)$ is some function of time to be determined.
Upon linearizing $\sigma$ the initial probability density (4.5)
becomes
\begin{equation}
P_{m_{0}}[m_{0}(\vec x)]=P_{0}[c m_{0}(\vec x)]
\end{equation}
where $c$ is a constant. Solving (4.15) by Fourier transform and
averaging over initial conditions with (4.16) one finds
\begin{equation}
C_{0}(\vec k,t)=S_{0}(t)L^{d}(t)g_{0} \bigl(kL(t) \bigr)
\end{equation}
\begin{equation}
G_{0}(\vec u,t)=S_{0}(t) \gamma \bigl(|\vec u|/L(t) \bigr)
\end{equation}
with
\begin{equation}
\left \{ \begin{array}{ll}
         L(t)=t^{\frac{1}{2}} \\
         \\
         g_{0}(kL)=\exp (-2(kL)^{2}) \\
         \\
         \gamma(u/L)=\exp (-\frac{u^{2}}{8 L^{2}}) \\
         \\
         S_{0}(t)=\frac{\Delta}{c^{2}L^{d}} \exp (2b(t)) \\
         \\
         b(t)=\int_{0}^{t} d{t^{\prime}} a(t^{\prime}).
         \end{array}
\right.
\end{equation}

\noindent Inserting the above expression
for $\gamma$ in (4.14) one obtains the
Ohta-Jasnow-Kawasaki$^{6}$ result for the scaling function which correctly
reproduces the behaviours (3.7) (3.8). It is then matter of studying
the behaviour of $S(t)$ and for this we must go over to the specific
implementations of the method.

\vspace{0.5cm}

{\bf \it On site linearization}

\vspace{0.5cm}

\noindent Making a direct extension to the field theory case of the procedure
adopted for one particle, let us look for a transformation $\sigma$
which linearizes the on site potential in (4.4)
\begin{equation}
-\frac{V^{\prime}(\sigma)}{\sigma^{\prime}(m)}=rm
\end{equation}
where $r$ is a constant. This yields
\begin{equation}
\frac{\partial m}{\partial t}=\nabla^{2}m+rm-Q(m)(\nabla m)^{2}
\end{equation}
where
\begin{equation}
Q(m)=-\frac{\sigma^{\prime \prime}(m)}{\sigma^{\prime}(m)}.
\end{equation}
With the double well potential of the form
\begin{equation}
V(\phi)=-\frac{r}{2} \phi^{2}+\frac{g}{4} \phi^{4}
\end{equation}
(4.20) reduces to (2.15) yielding as in (2.16)
\begin{equation}
\sigma(m)=\frac{m}{{[1+(g/r) m^{2}]}^{\frac{1}{2}}}
\end{equation}
and
\begin{equation}
Q(m)=3 \frac{g}{r} \frac{m}{[1+(g/r) m^{2}]}.
\end{equation}

\noindent However, contrary to what happens for one particle, even though the
transformation (4.24) manages to account for the saturation of
the order parameter, yet it is not sufficient to linearize the
equation of motion.
This is done by introducing an approximation, which is optimized
by the mean field prescription$^{11}$
\begin{equation}
Q(m)(\nabla m)^{2} \rightarrow 3 \frac{g}{r} <(\nabla m)^{2}>
< \frac{m^{2}}{1+(g/r) m^{2}} > \frac{m}{<m^{2}>}
\end{equation}
where averages must be computed self-consistenly. Note that
although (4.26) yields the best linear approximation to the equation
of motion, it remains an uncontrolled approximation since no small
parameter emerges which allows to compute corrections to it.
According to the general discussion made above the implementation
the GAF approximation requires, besides the
linearization of the equation of motion, also the linearization
of $\sigma$. Setting $\sigma(m)=m$ in (4.24), the initial
condition is given by (4.16) with $c=1$.

With (4.26) the equation of motion is of the form (4.15) with
\begin{equation}
a(t)=r-3 \frac{g}{r} D_{0}(t) \frac{S(t)}{S_{0}(t)}
\end{equation}
where
\begin{equation}
D_{0}(t)=<(\nabla m)^{2}>=\int_{\vec k} k^{2} C_{0}(\vec k,t).
\end{equation}
Next, using (4.13)
\begin{equation}
S(t)=\frac{r}{g} \bigl\{1-\sqrt{\frac{\pi r}{2gS_{0}(t)}}
e^{\frac{r}{2gS_{0}(t)}}
\bigl[1-erf \bigl( \sqrt{\frac{r}{2gS_{0}(t)}} \bigr) \bigr] \bigr\}
\end{equation}
and making the assumption to be verified a posteriori that $S_{0}(t)$
grows with time, asymptotically we have
\begin{equation}
S(t)=\frac{r}{g} \bigl\{1-\sqrt{\frac{\pi r}{2gS_{0}(t)}}+
O \bigl(\frac{1}{S_{0}(t)} \bigr) \bigr \}.
\end{equation}
Inserting into (4.27), to dominant order we get
\begin{equation}
{\dot b}(t)=r-3 \frac{D_{0}(t)}{S_{0}(t)}=r+O(t^{-1})
\end{equation}
which gives $b(t)=rt$. Next, using (4.19) we find
\begin{equation}
S_{0}(t)=\Delta \frac{\exp \{2rt \}}{t^{\frac{d}{2}}}
\end{equation}
which is consistent with the assumption made about $S_{0}(t)$. Finally,
inserting the above result into (4.30) we obtain that $S(t)$ saturates
exponentially fast to the equilibrium value $\phi_{eq}^{2}=r/g$, rather
than according to a power law as expected from (3.9).

\vspace{0.5cm}

{\bf \it KYG-theory}

\vspace{0.5cm}

\noindent The behaviour of $S(t)$ obtained above is what one finds resumming
the singular perturbation series of Kawasaki,Yalabik and Gunton$^{12}$ (KYG).
The KYG theory is contained in the above treatment as a particular case.
If in addition to the mean field approximation one makes also an expansion
in the nonlinear coupling $g$, to lowest order $Q(m) \equiv 0$ and the
equation of motion becomes
\begin{equation}
\frac{\partial m}{\partial t}=\nabla^{2} m+rm
\end{equation}
namely the auxiliary field coincides with the free field.
The transformation (4.24) together with (4.33) corresponds exactly
to the KYG theory, which therefore in the present context amounts
to the statement that all the important nonlinear features of the
problem are adequately taken care of by the transformation (4.24).

\vspace{0.5cm}

{\bf \it BH-theory}

\vspace{0.5cm}

\noindent If in (4.26) we keep the first order in $g$ the equation of motion
becomes
\begin{equation}
\frac{\partial m}{\partial t}=\nabla^{2} m
+[r-3 \frac{g}{r} <(\nabla m)^{2}>]m
\end{equation}
which is of the type of the equation obtained by Bray and Humayun$^{9}$ (BH)
starting from an {\it ad hoc} potential and which leads to the
correct behaviour for $S(t)$. In fact in this case (4.27) reduces to
\begin{equation}
{\dot b}(t)=r-3 \frac{g}{r} S_{0}(t) L^{d}(t)
\int_{k} k^{2} e^{-2 k^{2} t}
\end{equation}
and setting to zero the left hand side for large time
\begin{equation}
S_{0}(t) \sim L^{2}(t).
\end{equation}
Inserting this result in (4.30) the behaviour (3.9) of $S(t)$ is
recovered. This is due to the cancellation of $S_{0}(t)$ in the
denominator of (4.27), which occurs only in first order in $g$.
Notice that from the above result for $S_{0}(t)$ and the definition
(4.19) one obtains $a(t) \sim (d+2)/4t$ which coincides with the form
for $a(t)$ introduced by Oono and Puri$^{13}$ in their improvement
of the OJK theory.

\vspace{0.5cm}

In summary, the GAF approximation obtained via the linearization
of the on site potential i) does not describe correctly the saturation
law of the order parameter, except for the very special case where
the BH-theory applies, and ii) it is an uncontrolled approximation
since there is not a systematic expansion scheme within which it plays
the role of the zero order theory.

\vspace{0.5cm}

{\bf \it Mazenko transformation}

\vspace{0.5cm}

\noindent Let us now go to a different way of introducing the auxiliary
field due to Mazenko$^{4}$ where equation (4.2) is chosen in such a way
that $\sigma(m)$ reproduces the profile of the static interface
\begin{equation}
\sigma^{\prime \prime}(m)=V^{\prime}(\sigma).
\end{equation}
In this case $m(\vec x,t)$ has the physical interpretation of the
distance to the nearest interface. Using (4.37) in (4.4) we obtain
\begin{equation}
\frac{\partial m}{\partial t}=\nabla^{2} m+(1-(\nabla m)^{2})Q(m)
\end{equation}
where $Q(m)$ is still given by (4.22). Note that since
$V^{\prime}(\sigma)$
is an odd function from (4.37) and (4.22) follows
that also $Q(m)$ is an odd function. Thus, the mean field
linearization of (4.38) yields
\begin{equation}
\frac{\partial m}{\partial t}=\nabla^{2} m
+[1-<(\nabla m)^{2}>]H(t)m
\end{equation}
where $H(t)$ is some function of time whose explicit form is not
important. Hence Eq. (4.27) now gives
\begin{equation}
{\dot b}(t)=[1-S_{0}(t)L^{d} \int_{k} k^{2} e^{-2k^{2}t}]H(t)
\end{equation}
which, apart for the overall factor $H(t)$, is identical to (4.35)
and therefore leads to the same result (4.36) for $S_{0}(t)$ which
yields the correct behaviour of $S(t)$. Comparing (4.34) with (4.39)
we see that the BH-theory is a particular case arising with $H(t)$ constant.
Thus, the GAF approximation obtained within the static interface
approach yields correct results, but for the same reasons pointed
out above it remains an uncontrolled approximation.

\vspace{0.5cm}

{\bf \it Generalization of the transformation}

\vspace{0.5cm}

\noindent We end up this section by considering a generalization of the
transformation obtained by allowing for an explicit time dependence.
The idea is to see if in so doing one may get closer to the
linearization of the equation for $m(\vec x,t)$, as suggested
by recent work of Puri and Bray$^{14}$. Replacing (4.1) by
\begin{equation}
\phi(\vec x,t)=\sigma(t,m(\vec x,t))
\end{equation}
the equation for $m$
becomes
\begin{equation}
\frac{\partial m}{\partial t}=\nabla^{2}m+
\frac{1}{{\partial \sigma}/{\partial m}}
\bigl[\frac{\partial^{2} \sigma}{\partial m^{2}}(\nabla m)^{2}
-\frac{\partial \sigma}{\partial t}-V^{\prime}(\sigma) \bigr].
\end{equation}
Let us then determine the explicit dependence of $\sigma$ on $t$
by imposing
\begin{equation}
\frac{\partial \sigma}{\partial t}=-V^{\prime}(\sigma)
\end{equation}
which yields
\begin{equation}
\frac{\partial m}{\partial t}=\nabla^{2}m-Q(t,m)(\nabla m)^{2}
\end{equation}
with
\begin{equation}
Q(t,m)=-\frac{{\partial^{2} \sigma}/{\partial m^{2}}}
{{\partial \sigma}/{\partial m}}.
\end{equation}
Eq. (4.43) is nothing but the one particle equation of motion
which, using the potential (4.23), yields the solution (2.4) i.e.
\begin{equation}
\sigma(t,m)=\frac{\tau \sigma(0,m)}
{\bigl[1+\frac{g}{r}\sigma^{2}(0,m)(\tau^{2}-1) \bigr]^{\frac{1}{2}}}
\end{equation}
and
\begin{equation}
Q(t,m)=\frac{\frac{g}{r} \tau^2
\bigl[\sigma^{\prime \prime}\sigma^{2}
-3\sigma(\sigma^{\prime})^{2}\bigr]
+\bigl[\sigma^{\prime \prime}
-\frac{g}{r}
(\sigma^{\prime \prime}\sigma^{2}
-3\sigma(\sigma^{\prime})^{2}) \bigr]}
{\sigma^{\prime} \bigl[\frac{g}{r} \sigma^{2}
-\frac{g}{r} \tau^{2} \sigma^{2}-1 \bigr]}
\end{equation}
where the sigma's on the right hand side stand for $\sigma(0,m)$ and the
primes denote derivatives with respect to $m$.

Imposing
$\sigma^{\prime \prime}\sigma^{2}-3\sigma(\sigma^{\prime})^{2}=0$
we find $\sigma(0,m)=\pm (m)^{-1/2}$ and inserting into
(4.47) and (4.46) eventually we have
\begin{equation}
\sigma(t,m)=\pm \bigl[\frac{\tau^{2} m}
{1+(g/r)m(\tau^{2}-1)} \bigr]^{\frac{1}{2}}
\end{equation}
and
\begin{equation}
\frac{\partial m}{\partial t}=\nabla^{2}m
-\frac{3}{2}
\frac{\tau^{-2}}{\bigl[\tau^{-2}m-\frac{g}{r}(\tau^{-2}-1) \bigr]}
(\nabla m)^{2}.
\end{equation}
Indeed, the equation of motion for $m$ is "almost" linear
since the nonlinear
term vanishes exponentially fast, but the scheme it is not of
much use in generating a GAF approximation, since (4.48) cannot be
linearized.

\vspace{8mm}

\setcounter{chapter}{5}
\setcounter{equation}{0}

\section*{5 - Vector Fields}

\vspace{5mm}

Let us now consider the case of a vector order parameter with
$N$-components
${\vec \phi}(\vec x)=(\phi_{1}(\vec x),...,\phi_{N}(\vec x))$.
In this case a systematic expansion scheme about the GAF
approximation can be outlined, although its practical implementation
remains to be explored.

Phenomenological expectations in this case are a power law tail
in the scaling function of the structure factor$^{7,8}$
\begin{equation}
g(x) \sim x^{-(d+N)}
\end{equation}
which generalizes Porod's law and the saturation law$^{8}$
\begin{equation}
S(t)=\phi_{eq}^{2} \bigl[1-\frac{b}{L^{2}(t)}+O(L^{-3}) \bigr]
\end{equation}
in place of (3.9).
Considering the equation of motion
\begin{equation}
\frac{\partial \phi_{\alpha}(\vec x,t)}{\partial t}=
\nabla^{2}\phi_{\alpha}(\vec x,t)
-\frac{\partial}{\partial \phi_{\alpha}}V({\vec \phi}(\vec x,t))
\end{equation}
with the potential
\begin{equation}
V(\vec \phi)=-\frac{r}{2}{\vec \phi}^{2}
+\frac{g}{4N}({\vec \phi}^{2})^{2}
\end{equation}
the auxiliary field ${\vec m}(\vec x,t)$ is introduced by generalizing
to the vector case the transformation (4.24)
\begin{equation}
\sigma_{\alpha}(\vec m)=
\frac{m_{\alpha}}
{\bigl[1+\frac{g}{rN} {\vec m}^{2} \bigr]^{\frac{1}{2}}}
\end{equation}
which yields the equation of motion for $\vec m$
\begin{eqnarray}
&& \frac{\partial m_{\alpha}}{\partial t}=
\nabla^{2}m_{\alpha}+rm_{\alpha}-\frac{g}{rN}
\bigl \{ m_{\alpha} \sum_{\gamma}
( \nabla m_{\gamma})^{2}  \nonumber \\
&& \\
&& +\frac{\bigl[2 \nabla m_{\alpha} \cdot
\sum_{\gamma} (m_{\gamma} \nabla m_{\gamma})
-\frac{g}{rN}m_{\alpha}(m_{\gamma} \nabla m_{\gamma})^{2} \bigr]}
{\bigl[1+\frac{g}{rN} \sum_{\beta} m_{\beta}^{2} \bigr]} \bigr \}. \nonumber
\end{eqnarray}
For the equal time correlation function
$G(\vec u,t)
=<\phi_{\alpha}({\vec x}_{1},t)\phi_{\alpha}({\vec x}_{2},t)>$,
which is independent of $\alpha$ due to the rotational symmetry of
the potential, we have
\begin{equation}
G(\vec u,t)=\int d{\vec m}_{1} d{\vec m}_{2}
P_{m}({\vec m}_{1},{\vec x}_{1} t;{\vec m}_{2},{\vec x}_{2} t)
\sigma_{\alpha}({\vec m}_{1}) \sigma_{\alpha}({\vec m}_{2})
\end{equation}
where $P_{m}$, which is related to the initial probability density
$P_{0}$ through the analogues of (4.7) and (4.9), depends explicitely
on $N$ through $\sigma$ and $h$.
As previously stated $P_{m}$ becomes gaussian upon linearizing
$\sigma$ and $h$. We now show that this is achieved by taking the
large-$N$ limit.
The major difference with the scalar case than
is that now there emerges $\lambda=1/N$ as the natural parameter  which
yields the gaussian approximation in the limit
$\lambda \rightarrow 0$.

Taking the limit $N \rightarrow \infty$ terms of the type
$\frac{1}{N} \sum_{\alpha} q_{\alpha}$ in (5.5) and (5.6)
are replaced by the average $<q_{\alpha}>$ yielding the linear
equations
\begin{equation}
\sigma_{\alpha}(\vec m)=
\frac{m_{\alpha}}{\bigl[1+\frac{g}{r}S_{0}(t) \bigl]^{\frac{1}{2}}}
\end{equation}
and
\begin{equation}
\frac{\partial m_{\alpha}}{\partial t}=
\nabla^{2}m_{\alpha}+\bigl[r-\frac{g}{r}
<(\nabla m_{\alpha})^{2}> \bigr]m_{\alpha}
\end{equation}
since $<m_{\gamma} {\vec \nabla}m_{\gamma}>$ vanishes.
Hence, as anticipated, in the large-$N$ limit the auxiliary
field $m$ is gaussian. Furthermore Eq. (5.9) is of the BH-type
yielding (4.36) for $S_{0}(t)=<m_{\alpha}^{2}(t)>$.

It is important to realize that the large-$N$ limit we are
considering here is quite different from the
usual large-$N$ limit$^{15}$ performed on the equation of motion
(5.3) for $\vec \phi$. The latter one is recovered in the
present context by taking the large-$N$ limit, namely
using (5.8), also in the explicit $\sigma_{\alpha}$'s
appearing in (5.7) and eventually obtaining
\begin{equation}
G(\vec u,t)=\frac{r}{g} \bigl[1-\frac{r}{gL^{2}} \bigr]
\exp (-\frac{u}{8L^{2}}).
\end{equation}
Instead, according to the general structure of the GAF
approximation which we have repeatedly illustrated above,
$N$ must be kept fixed to whatever value it has been
originally specified in the explicit $\sigma_{\alpha}$'s
in (5.7), while the $N \rightarrow \infty$ limit is taken
inside $P_{m}$. In so doing from (5.5) and (5.7) we obtain
the Bray, Puri and Toyoki $^{7}$ (BPT) result for the scaling function
\begin{eqnarray}
&& f \bigl(\frac{u}{L(t)})=<\hat{m}({\vec x}_{1},t) \cdot
\hat{m}({\vec x}_{2},t)> \nonumber \\
&& \\
&& =\frac{N \gamma}{2 \pi}
\left [ B \bigl(\frac{N+1}{2},\frac{1}{2} \bigr)^{2}
F \bigl(\frac{1}{2},\frac{1}{2};\frac{N+2}{2},\gamma^{2}
\bigr) \right ] \nonumber
\end{eqnarray}
where$B(x,y)$ is the beta function,
$F(a,b;c;z)$ the hypergeometric function and $\gamma(u/L)$
is given by (4.19). From the above result follows the power law tail (5.1).
Furthermore, from $S(t)=<\sigma_{\alpha}^{2}>$ we obtain
\begin{equation}
S(t)=\frac{1}{N} \int \frac{d \vec{m}}{(2\pi S_{0})^{N/2}}
\frac{m^{2}}{\bigl [1+\frac{gm^{2}}{rN} \bigr ]} e^{-\frac{m^{2}}
{2S_{0}}}
\end{equation}
and carrying out the integral
\begin{equation}
S(t)=\frac{Nr}{2g} \bigl (\frac{Nr}{2gS_{0}} \bigr )^{N/2}
e^{\frac{Nr}{2gS_{0}}} \Gamma (-\frac{N}{2},\frac{Nr}{2gS_{0}})
\end{equation}
where $\Gamma (x,y)$ is the incomplete gamma function.
Expanding up to first order in $1/S_{0}$ we obtain
\begin{equation}
S(t) \sim \left \{ \begin{array}{ll} \phi_{eq}^{2}
\bigl[1-\frac{N}{(N-2)}\frac{r}{gS_{0}(t)} \bigr] & \mbox{for $N>2$} \\
\\
\phi_{eq}^{2} \bigl [1+\frac{r}{2gS_{0}(t)} \bigr ] & \mbox{for $N=2$}
\end{array}
\right.
\end{equation}
which yields the power law behaviour (5.2), contrary to the
exponential saturation which one obtains in the BTP approach.

These results show that the expected phenomenological
behaviour is obtained at zero order within the
$1/N$-expansion of the probability density
of the auxiliary field. In principle, systematic
corrections could be obtained via the higher order terms
in the $1/N$-expansion of $P_{m}$, although
we do not expect that such a scheme of computation
might be easily implemented in practice.
It is worth pointing out that the scheme for the
systematic improvement of the GAF approximation
for vector fields presented here is conceptually different
from that proposed by BH in two respects:
i) while we use the standard quartic potential (5.4)
BH need to invoke an {\it ad hoc} potential which cannot
even be written in closed form and ii) the expansion is
made in $1/N$ where here $N$ is the number of components of the
order parameter rather than the number of components of an
additional internal color index.
Finally, the comparison between (5.10) on one side and (5.11) (5.14)
on the other clearly shows the difference between the standard
$1/N$-expansion and the one we have presented here. The most
important point is that while there are no localized
defects in lowest order in the usual $1/N$-expansion since
the correlation function (5.10) decays exponentially, the power
law tail (5.1) implied by (5.11) shows that our reformulation of
the $1/N$-expansion describes defects in lowest order.

\vspace{8mm}

\setcounter{chapter}{6}
\setcounter{equation}{0}

\section*{6 - Concluding remarks}

\vspace{5mm}

In conclusion, in this paper we have analysed the sequence
of steps which must be taken within the framework of a first
principles theory in order to generate GAF approximations.
The analysis has been restricted to systems with non conserved
order parameter. The idea was to look for the systematic
expansion scheme which allows to control the corrections to
the GAF approximations. A project of this type is suggested
by the physical motivation behind the introduction of the
auxiliary field. This being more smooth and less non linear
than the order parameter field, hopefully should be tractable
in perturbation theory. Our results are negative for the
scalar case, in the sense that we are unable to come up with
the expansion scheme within which the GAF approximation can be
identified with the zero order approximation.
It should be mentioned that there are indications$^{16,9}$
that the GAF approximation becomes exact in the limit of
infinite space dimensionality, suggesting the $1/d$-expansion
as a possible systematic expansion scheme. This is an interesting
line of research worth to be further investigated.

The outlook is
somewhat better in the case of a vector order parameter.
In this case one can set up the theory in such a way that the
large-$N$ limit yields the GAF approximation. Consequently one
can expect that there exists a $1/N$-expansion where corrections
to the GAF approximation are generated systematically.
Finally, approximations which go beyond the GAF approximation
have been introduced recently by Mazenko$^{17}$. In future work
we plan to look for the connection between that work and the
point of view developed here.

\newpage

{}~~\\
{}~~\\
{\bf  References}
{}~~\\
\begin{enumerate}

\item J.D.Gunton, M.San Miguel and P.S.Sahni in {\it Phase
transitions and critical phenomena} edited by C.Domb and J.L.Lebowitz
(Academic Press, New York 1983) Vol.8, p.267

\noindent H.Furukawa, {\it Adv. Phys.} {\bf 34}, 703, (1985)

\noindent K.Binder {\it Rep. Progr. Phys.} {\bf 50}, 783, (1987)

\item A.J.Bray in {\it Phase transitions in systems with competing
energy scales} edited by T.Riste and D.Sherrington (Kluwer
Academic, Boston 1993)

\item A.J.Bray and K.Humayun, {\it Phys. Rev. Lett.} {\bf 68},
1559, (1992)

\item G.F.Mazenko, {\it Phys. Rev. B} {\bf 42}, 4487, (1990) and
{\it ibidem} {\bf 43}, 5747, (1991)

\item C.Yeung, Y.Oono and A.Shinozaki, {\it Phys. Rev. E}
{\bf 49}, 2693, (1994)

\item T.Ohta, D.Jasnow and K.Kawasaki, {\it Phys. Rev. Lett.}
{\bf 49}, 1223, (1982)

\item A.J.Bray and S.Puri, {\it Phys. Rev. Lett.} {\bf 67},
2670, (1991)

\noindent H.Toyoki, {\it Phys. Rev. B} {\bf 45}, 1965, (1992)

\item F.Liu and G.F.Mazenko, {\it Phys. Rev. B} {\bf45}, 6989,
(1992)

\item A.J.Bray and K.Humayun, {\it Phys. Rev. E} {\bf 48}, R1609,
(1993)

\item M.Suzuki, {\it Progr. Theor. Phys.} {\bf 56}, 77, (1976);
{\it J. Stat. Phys.} {\bf 16}, 11, (1977); {\it Phys. Lett.} {\bf 67A},
339, (1978)

\noindent F.de Pasquale andP.Tombesi, {\it Phys. Lett.} {\bf 72A},
7,(1979)

\item M.C.Valsakumar, K.P.N.Murthy and G.Ananthakrishna,
{\it J. Stat. Phys.} {\bf 30}, 617, (1983)

\item K.Kawasaki, M.C.Yalabik and J.D.Gunton, {\it Phys. Rev. A}
{\bf 17},455, (1978)

\item Y.Oono and S.Puri, {\it Mod. Phys. Lett. B} {\bf 2}, 861,
(1988)

\item S.Puri and A.J.Bray, {\it J. Phys. A: Math. Gen.} {\bf 27}, 453,
(1994)

\item A.Coniglio and M.Zannetti, {\it Europhys. Lett.} {\bf 10},
575, (1989)

\item F.Liu and G.F.Mazenko, {\it Phys. Rev. B} {\bf 45}, 4656, (1992)

\item G.F.Mazenko, {\it Post gaussian approximations in phase
ordering kinetics}, preprint, 1994

\end{enumerate}

\end{document}